\documentclass[a4paper,UKenglish]{lipics-v2021}

\usepackage{xspace}
\usepackage{thm-restate}
\usepackage[textsize=small]{todonotes}
\usepackage{stmaryrd}

\usepackage{graphicx} %
\usepackage{tikz} 
\usetikzlibrary{arrows, automata, positioning,calc}
\usetikzlibrary{shapes,snakes}

\usepackage{tikz} 
\usetikzlibrary{arrows, automata, positioning,calc}
\usetikzlibrary{shapes,snakes}

\usepackage{floatrow}
\newfloatcommand{capbtabbox}{table}[][\FBwidth]

\newcommand{\Oo}{\mathcal{O}}
\newcommand{\D}{\mathbb{D}}

\newcommand{\A}{\mathcal{A}}
\newcommand{\B}{\mathcal{B}}

\newcommand{\N}{\mathbb{N}}

\newcommand{\Q}{\mathbb{Q}}

\newcommand{\pref}{\textup{pref}}

\newcommand{\init}{\textup{init}}

\newcommand{\fresh}{\textup{fresh}}

\newcommand{\ds}{\textup{data}}%

\DeclareSymbolFont{extraup}{U}{zavm}{m}{n}
\DeclareMathSymbol{\varheart}{\mathalpha}{extraup}{86} %
\DeclareMathSymbol{\vardiamond}{\mathalpha}{extraup}{87} %

\newcommand\mN{\mathbb N}

\newcommand\ie{\emph{i.e.}}
\newcommand\eg{\emph{e.g.}}

\newcommand{\domain}{\mathbb{D}}
\newcommand{\data}{\ds}

\newcommand{\kqhide}[1]{\textcolor{teal}{sth old is hidden here}}

\newcommand{\locs}{\mathcal{L}}
\newcommand{\loc}{\ell}
\newcommand{\edges}{E}
\newcommand{\true}{\texttt{true}}
\newcommand{\acc}{\textup{acc}}

\newcommand{\config}{C}
\newcommand{\sconfig}{S} %

\newcommand{\sNodes}{\mathbb{S}} %
\newcommand{\sTo}{\Rightarrow} %

\renewcommand{\succ}{\textup{Succ}} 
\newcommand{\vect}[1]{\boldsymbol{#1}}

\newcommand{\indiscernible}[3]{{#1} \equiv_{#3} {#2}}

\newcommand{\register}{r}
\newcommand{\registers}{\mathcal{R}}

\newcommand{\URA}{\textup{URA}}

\newcommand{\currentinput}{\#}

\newcommand{\supp}{\textup{supp}}

\newcommand{\GURA}{\textup{GURA}}
\newcommand{\GRA}{\textup{GRA}}

\newcommand{\RA}{\textup{RA}}

\renewcommand{\frac}{\textup{frac}}

\newcommand{\nctwo}{\textup{NC$^2$}\xspace}
\newcommand{\ptime}{\textup{PTime}\xspace}
\newcommand{\pspace}{\textup{PSpace}\xspace}
\newcommand{\npspace}{\textup{NPSpace}\xspace}
\newcommand{\exptime}{\textup{ExpTime}\xspace}
\newcommand{\expspace}{\textup{ExpSpace}\xspace}
\newcommand{\twoexptime}{\textup{2-ExpTime}\xspace}
\newcommand{\twoexpspace}{\textup{2-ExpSpace}\xspace}

\newcommand\EXPSPACE{\expspace}

\newcommand{\clean}{pruned\xspace} %
\newcommand{\const}{M} %
\newcommand{\fatconst}{\mathbf{M}} %

\title{New Techniques for Universality in Unambiguous Register Automata}

\author{Wojciech Czerwi\'nski}{University of Warsaw}{wczerwin@mimuw.edu.pl}{0000-0002-6169-868X}{Supported by the European Research Council (ERC) grant LIPA, grant agreement No 683080.}
\author{Antoine Mottet}{Department of Algebra, Faculty of Mathematics and Physics, Charles University in Prague\and\url{http://www.karlin.mff.cuni.cz/~mottet/}}{mottet@karlin.mff.cuni.cz}{0000-0002-3517-1745}{This author has received funding from the ERC under the European Union's Horizon 2020 research and innovation programme (grant agreement No 771005).}
\author{Karin Quaas}{University of Leipzig}{url}{orcid}{Funded by the Deutsche Forschungsgemeinschaft (DFG), project 406907430.}
\Copyright{Wojciech Czerwi\'nski, Antoine Mottet, Karin Quaas}
\authorrunning{W. Czerwi\'nski, A. Mottet, K. Quaas}


\acknowledgements{We thank Lorenzo Clemente, S{\l}awomir Lasota, and Rados{\l}aw Pi{\'o}rkowski 
for inspiring discussions on $\URA$.}
\relatedversion{}

\EventEditors{John Q. Open and Joan R. Access}
\EventNoEds{2}
\EventLongTitle{42nd Conference on Very Important Topics (CVIT 2016)}
\EventShortTitle{CVIT 2016}
\EventAcronym{CVIT}
\EventYear{2016}
\EventDate{December 24--27, 2016}
\EventLocation{Little Whinging, United Kingdom}
\EventLogo{}
\SeriesVolume{42}
\ArticleNo{23}

\hideLIPIcs
\nolinenumbers

\begin{document}

\maketitle

\begin{abstract}
Register automata are finite automata equipped with a finite set of registers ranging over the domain of some relational structure like $(\N;=)$ or $(\Q;<)$. 
Register automata process words over the domain, 
and along a run of the automaton, 
the registers can store data from the input word for later comparisons. 
It is long known that the universality problem, i.e., the problem to decide whether a given register automaton accepts all words over the domain, 
is undecidable. %
Recently, we proved the problem to be decidable in 2-ExpSpace if the register automaton under study is over $(\N;=)$ and unambiguous, i.e., every input word has at most one accepting run; this result was shortly after improved to 2-ExpTime by Barloy and Clemente. In this paper, we go one step further and 
prove that the problem is in ExpSpace, and in PSpace if the number of registers is fixed. Our proof is based on new techniques that additionally allow us to show that 
the problem is in PSpace for single-register automata over $(\Q;<)$.  
As a third technical contribution we prove that the problem is decidable (in ExpSpace) for a more expressive model of unambiguous register automata, where the registers can take values nondeterministically, if defined over $(\N;=)$ and  only one register is used. 
\end{abstract}

\section{Introduction}
Certainly, determinism plays a central role in the research about computation models. Recently, a lot of active research work~\cite{BarloyClemente,DBLP:conf/concur/CzerwinskiFH20,Bostan2020,Paul2020,DBLP:conf/stacs/MottetQ19} is devoted
to its weaker form: \emph{unambiguity}. 
A system is \emph{unambiguous} if for every input word there is at most one accepting run.
Unambiguous systems exhibit elegant properties; in particular many natural computational problems turn out to be easier
 compared to the general case. 
A prominent example is the \emph{universality problem} for finite automata, \ie, the problem of deciding whether a given automaton accepts \emph{every} input word. 
It  is in \ptime~\cite{DBLP:journals/siamcomp/StearnsH85}
and even in \nctwo~\cite{DBLP:journals/ipl/Tzeng96} in the unambiguous case, as opposed to \pspace-completeness in the general case.

In his seminal overview article about unambiguity, 
Colcombet~\cite{DBLP:conf/dcfs/Colcombet15} states some very natural conjectures about unambiguous systems that are so fundamental that
one can be surprised that they are still open. 
An example conjecture, motivated by the fact that the universality problem
for unambiguous finite automata is in \ptime, was that for every unambiguous finite automaton
the complement of its language can be accepted 
by another unambiguous finite automaton with at most polynomial size with respect to the size of the original automaton. 
This conjecture
was surprisingly resolved negatively by Raskin~\cite{DBLP:conf/icalp/Raskin18}, who provided a family of automata where a blowup
$\Theta(n^{\log\log\log(n)))})$ is unavoidable. Still, 
a lot of other natural questions remain unresolved. Some of them are not algorithmic (as the above one), while others
ask for the existence of faster algorithms in the unambiguous case.

Usually one cannot hope for designing more efficient algorithms for the emptiness problem,
as it is often easy to transform a nondeterministic system to a deterministic (and thus unambiguous) 
system which has empty language if and only if the accepted language of the original system is empty. 
Indeed, it is often sufficient to change the labelling of 
every transition of the system to its unique transition name. This transformation preserves the emptiness property, but not much more. 
Therefore there is a hope that the unambiguity assumption may result in faster solving of problems like universality, equivalence and language containment. %
Recently there was a substantial amount of research in this area~\cite{DBLP:journals/ipl/IsaakL12,DBLP:conf/lata/BousquetL10,DBLP:conf/icalp/DaviaudJLMP018,DBLP:conf/stacs/MottetQ19,DBLP:conf/concur/CzerwinskiFH20,BarloyClemente}. 
The considered problem is often the universality problem. 
Indeed, the universality problem is probably the easiest nontrivial problem for which there is a hope to obtain an improvement in the
unambiguous case. 
Equivalence and containment are often not much harder, 
even though sometimes a bit more involved techniques are needed.

For \emph{register automata}, 
this line of research was started in~\cite{DBLP:conf/stacs/MottetQ19}. 
Register automata ($\RA$, for short) extend finite automata with a finite set of registers that take values from an infinite data domain for later comparisons. 
More detailed, $\RA$  are defined over a relational structure, like $(\N;=)$ or $(\Q;<,=)$; they process finite words over the domain of the relational structure, and the registers can store values from the input word for comparing them using the relations provided by the relational structure. 
In the more expressive model of register automata \emph{with guessing} ($\GRA$)  the registers can even take arbitrary values. 
In~\cite{DBLP:conf/stacs/MottetQ19} it is shown that for unambiguous $\RA$ ($\URA$) over $(\N;=)$ the containment problem is in \twoexpspace and
in \expspace for a fixed number of registers.
Without the unambiguity assumption, 
this problem is known to be much harder.
Concretely, the universality problem is undecidable as soon as the automaton uses two registers~\cite{DBLP:journals/tcs/KaminskiF94,DBLP:journals/tocl/NevenSV04, DBLP:conf/fossacs/DemriLS08},  and Ackermann-complete in the one-register case~\cite{DBLP:conf/lics/FigueiraFSS11}. 
In the case of $\GRA$ even the one-register case is undecidable.\footnote{A proof for undecidability can be done using a reduction from the undecidable reachability problem for Minsky machines, following the lines of the proof of Theorem 5.2 in~\cite{DBLP:journals/tocl/DemriL09}. The nondeterministic guessing can be used to express that there exists some decrement for which there is no matching preceding increment.}
The result for $\URA$ in~\cite{DBLP:conf/stacs/MottetQ19} was improved by Barloy and Clemente~\cite{BarloyClemente} who have shown that the problem is in \twoexptime
and in \exptime for a fixed number of registers, using very different tools such as linear recursive sequences in two dimensions.

\paragraph*{Our contribution}
Our result improves statements of Barloy and Clemente~\cite{BarloyClemente} even further. 
We provide three results shown by two different techniques.
Our first technique is to show that in some cases one can assume that only a linear or exponential number of different configurations
can be reached via an input word.
This claim immediately provides an improved upper bound compared to~\cite{BarloyClemente}.

\begin{restatable}{theorem}{urabound}\label{thm:urabound}
The containment problem $L(\A)\subseteq L(\B)$ is in \expspace, if $\A$ is an $\RA$ and $\B$ is a $\URA$ over $(\mathbb N;=)$.
The containment problem is in $\pspace$ on inputs $\A,\B$ both having a bounded number of registers.
\end{restatable}

 This approach can also be applied to unambiguous one-register automata over $(\mathbb Q;<,=)$.
\begin{restatable}{theorem}{uraorderbound}\label{thm:uraorderbound} 
The universality problem for one-register $\URA$ over $(\Q;<,=)$ is in $\pspace$. 
\end{restatable}
However, 
we will see that the techniques for $\URA$ do not work for unambiguous $\GRA$ ($\GURA$), not even in the one-register case.  
In that case we solve the universality problem, and even the containment  problem, with the use of more sophisticated analysis. In short, we show that we can modify the set of reachable configurations
such that it becomes small and equivalent in some sense, which also allows us to obtain a more efficient algorithm.
\begin{restatable}{theorem}{gurabound}
\label{thm_gura}
The containment problem $L(\A)\subseteq L(\B)$ is in $\textup{\EXPSPACE}$, if $\A$ is a $\GRA$ over $(\mathbb N;=)$ and $\B$ is a one-register $\GURA$ over $(\mathbb N;=)$. 
\end{restatable}

We recently learned that, independently from our work, Bojańczyk, Klin and Moerman claim a yet unpublished
result about orbit-finite vector spaces, which implies an \exptime algorithm for $\GURA$ and \pspace complexity if the number of registers is fixed.
However, we believe that our contribution does not only provide an improved complexity of the considered problem,
but also techniques that can be useful in future research on unambiguous systems. %

\section{Preliminaries}
In this section, 
we define \emph{register automata}, introduced by Kaminski et al~\cite{DBLP:journals/tcs/KaminskiF94,KaminskiZeitlin}. 
We start with some basic notions used throughout the paper. 
We use $\Sigma$ to denote a finite alphabet, 
and $\N$ and $\Q$ denote the set of non-negative integers and rational numbers, respectively. 
Given $a,b\in\N$ with $a\leq b$, we write $[a,b]$ to denote the set $\{a,a+1,\dots,b\}$.

A \emph{relational structure} is a tuple $\mathcal{D}=(\domain; R_1,\dots, R_k)$, where 
$\domain$ is an infinite domain, and $R_1,\dots,R_k$ are binary relations over $\domain$, and we assume that $R_k$ is the equality relation. 
In this paper, we are mainly interested in the relational structures 
$(\N;=)$ of the non-negative integers with equality,  
and $(\Q;<,=)$ of the rationals with the usual order and equality relations. %

A \emph{data word} is a finite sequence $(\sigma_1,d_1)\dots(\sigma_k,d_k) \in (\Sigma\times\domain)^*$. If $\Sigma=\{\sigma\}$ is a singleton set,  
we may write $d_1 \cdot d_2\cdot \, \dots\, \cdot d_k$ shortly for $(\sigma,d_1)(\sigma,d_2)\dots(\sigma,d_k)$. 
We use $\varepsilon$ to denote the empty data word.  
A \emph{data language} is a set of data words. 
We use $\data(w)$ to denote the set $\{d_1,\dots,d_k\}$ of all data occurring in $w$.

Let $\domain_\bot$ denote the set $\domain\cup\{\bot\}$, where $\bot\not\in\domain$. 
We let $\bot \neq d$ for all $d\in\domain$, and $\bot$ is incomparable with respect to $\leq$ to all $d\in\domain$. 
We use boldface lower-case letters like $\vect{a}, \vect{b}, \dots, \vect{u} \dots$ to denote tuples in $\domain_\bot^n$, where $n\in\mN$.
Given a tuple $\vect{a}\in\domain_\bot^n$, we write $a_i$ for its $i$-th component, and $\data(\vect{a})$ denotes the set $\{a_1,\dots,a_n\}\subseteq\domain_\bot$ of all data  occurring in $\vect{a}$.

Let $\registers=\{\register_1,\dots,\register_n\}$ be a finite set of \emph{registers}. 
A \emph{register valuation} is a mapping $\vect{u}:\registers\to\domain_\bot$; we may write $u_i$ as shorthand for $\vect{u}(\register_i)$. Let $\domain_\bot^\registers$ denote the set of all register valuations. 
A \emph{register constraint over $\mathcal{D}$ and $\registers$} is defined by the grammar
\begin{align*}
\phi ::= \true \,\mid\,  \, R(t_1,t_2) \, \,\mid\,  \neg\phi \,\mid\, \phi \wedge \phi 
\end{align*} 
where $R$ is a binary relation symbol from the relational structure $\mathcal{D}$, 
and $t_i\in \{\currentinput\}\cup\{\register,\dot\register\mid\register\in\registers\}$. 
Here $\currentinput$ is a symbol representing the current input datum, $\register$ refers to the current value of the register $\register$, and $\dot\register$ refers to the future value of the register $\register$.  
We use $\Phi(\mathcal{D},\registers)$ to denote the set of all register constraints over $\mathcal{D}$ and $\registers$.  
The satisfaction relation $\models$ on $\domain_\bot^\registers\times\domain\times\domain_\bot^\registers$ is defined by structural induction as follows. We only give some atomic cases; the other cases can be derived easily. We have  $(\vect{u},d,\vect{v}) \models \phi$ if 
\begin{itemize}
\item $\phi$ is of the form $\true$,
\item $\phi$ is of the form $R(\register_i,\currentinput)$ and $\mathcal D\models R(u_i,d)$, 
\item $\phi$ is of the form $R(\dot\register_i,\register_i)$  and $\mathcal D\models R(v_i,u_i)$,  
\item $\phi$ is of the form $R(\dot\register_i,\currentinput)$ and $\mathcal D\models R(v_i,d)$. 
\end{itemize}
For example, 
$\phi := \neg(\register=\currentinput)\wedge(\dot\register=\register)$ is a register constraint over $(\N;=)$ and $\registers=\{\register\}$, and we have $(1,2,1)\models \phi$, whereas $(1,2,3)\not\models\phi$.  

It is important to note that only register constraints of the form $\dot\register=\register$ and $\dot\register=\currentinput$ 
uniquely determine the new value of $\register$. 
In absence of such a register constraint, the register
$\register$ can nondeterministically take any of infinitely many data values from $\domain$, with the following restrictions: 
the register constraint $\neg(\dot\register=\currentinput)$ requires that the new value of $\register$ is different from the current input datum, so that $\register$ may take any  datum in $\domain$ except for the input datum.  Likewise, the register constraint $\neg(\dot\register=\register)$ requires that $\register$ takes any datum in $\domain$ except for the current value of $\register$. 
Register automata that allow for such nondeterministic \emph{guessing} of future register values are also called \emph{register automata with guessing}. 
Formally, a register automaton with guessing (\GRA) over $\mathcal{D}$ and $\Sigma$ is a tuple $\A = (\registers,\locs, \loc_{\init}, \locs_{\acc}, \edges)$, where
\begin{itemize}
\item $\registers$ is a finite set of registers, 
\item $\locs$ is a finite set of locations, 
\item $\loc_{\init}\in\locs$ is the initial location, 
\item $\locs_{\acc}\subseteq\locs$ is the set of accepting locations, 
\item $\edges\subseteq\locs\times\Sigma\times\Phi(\mathcal{D},\registers)\times\locs$ is a finite set of edges.
\end{itemize}
If every edge of $\A$ contains some constraint of the form $\dot\register=\register$ or $\dot\register=\currentinput$, for every $\register\in\registers$, so that the future value of every register is uniquely determined, then we simply speak of \emph{register automata} (\RA, for short), \ie, register automata without guessing. 
If the number of registers of a $\GRA$ ($\RA$, respectively) is fixed to $k\in\N$, then we speak of $k$-$\GRA$ ($k$-$\RA$, respectively).

A \emph{state} of $\A$  is a pair $(\loc,\vect{u})\in \locs\times\domain_\bot^\registers$, where $\loc$ is the current location and $\vect{u}$ is the current register valuation.  
Abusing notation a bit, we usually write $\loc(\vect{u})$ instead of $(\loc,\vect{u})$. 
The state $\loc_\init(\vect{u}_\init)$, where $\vect{u}_\init$ maps every register $r\in\registers$ to $\bot$, 
is called the \emph{initial state}, and 
a state $\loc(\vect{u})$ is called \emph{accepting} if $\loc\in\locs_\acc$. 
Given two states $\loc(\vect{u})$ and $\loc'(\vect{u'})$ and some input letter $(\sigma,d)\in\Sigma\times\domain$, we postulate a transition $\loc(\vect{u})\xrightarrow{\sigma,d}_\A\loc'(\vect{u'})$ if there exists some edge $(\loc,\sigma,\phi,\loc')\in\edges$ such that $(\vect{u},d,\vect{u'})\models\phi$. 
A \emph{run} of $\A$ on the data word 
$(\sigma_1,d_1)\dots(\sigma_k,d_k)$ is a  sequence 
$\loc_0(\vect{u^0}) \xrightarrow{\sigma_1,d_1}_\A \loc_1(\vect{u^1}) \xrightarrow{\sigma_2,d_2}_\A \dots \xrightarrow{\sigma_k,d_k}_\A \loc_k(\vect{u^k})$ of such transitions. 
We say that a run as above \emph{starts in $\loc_0(\vect{u^0})$};
similarly, the run \emph{ends in $\loc_k(\vect{u^k})$}. 
A state $\loc(\vect{u})$ is \emph{reachable} in $\A$ if there exists a run that ends in $\loc(\vect{u})$. 
A run is \emph{initialized} if it starts in the initial state,  
and a run 
is \emph{accepting} if it ends in some accepting state.  
A data word $w$ is \emph{accepted from $\loc(\vect{u})$} if there exists an accepting run on $w$ that starts in $\loc(\vect{u})$. 
The data language \emph{accepted} by $\A$, denoted by $L(\A)$, is the set of data words that are accepted from the initial state.  

A $\GRA$ is \emph{unambiguous} if for every input data word $w$ there is at most one initialized accepting run. Note that unambiguity is a semantic condition; it can be checked in polynomial time~\cite{DBLP:conf/dcfs/Colcombet15}. We write $\GURA$ and $\URA$ to denote unambiguous $\GRA$ and $\RA$, respectively.

\begin{example}
Let us study the behaviour of the $1$-$\GRA$ depicted  in Figure \ref{fig:gura}.  
The $\GRA$ is over $(\N;=)$ and the singleton alphabet $\Sigma=\{\sigma\}$ (we omit the letter $\sigma$ from all transitions in the figure). 
Suppose the first input letter is $d_1$.  
In order to satisfy the constraint of the transition from $\loc_1$ to $\loc_2$, the automaton has to nondeterministically guess some datum $d'\neq d_1$ and store it into its register $\register$. 
Being in the state $\loc_1(d')$, the automaton can only move to the accepting location $\loc_2$ if the next input datum is equal to $d'$ (indicated by the constraint $\register=d$); 
for every other input letter, 
the automaton satisfies the constraint $\neg(\register=d)$ and stays in $\loc_1$, and it keeps the register value to satisfy the constraint $\dot\register=\register$. 
In this way, the automaton accepts the language $\{d_1 \cdot \, \dots \, \cdot d_k\mid \forall  k\geq 2 \, \,  \forall 1\leq i<k. \, \,  d_i\neq d_k\}$. 
Note that the automaton is unambiguous: for every input data word there %
is only one accepting run. 
We remark that the accepted data language cannot be accepted by any $\RA$ (without guessing)~\cite{KaminskiZeitlin}. Hence, $\GRA$ are more expressive than $\RA$. 
\end{example}

\begin{figure}
\begin{floatrow}
\ffigbox{%
  \scalebox{.9}{ %
\begin{tikzpicture}[->,>=stealth',shorten >=1pt,auto,node distance=4cm,thick,node/.style={circle,draw,scale=0.9}, roundnode/.style={circle, draw=black, thick, minimum size=6mm},]
\tikzset{every state/.style={minimum size=0pt}};
\node[roundnode,initial, initial text={}]  (1) at (0,0) {$\loc_0$};
\node[roundnode]  (2) at (2,0) {$\loc_1$};
\node[roundnode,accepting]  (3) at (4,0) {$\loc_2$};
\path [->] (1) edge node[above] {\scriptsize{$\neg(\dot\register=\currentinput)$}} (2);

\path [->] (2) edge node[above] {\scriptsize{$\dot\register=\register$}} 
node [above,yshift=3mm] {\scriptsize{$\register=\currentinput$}}
(3);

\path [->] (2) edge [loop above] node[above] {\scriptsize{$\dot\register=\register$}}
node [above,yshift=3mm] {\scriptsize{$\neg(\register=\currentinput)$}} (2);
 	\end{tikzpicture}  
  }
}{%
  \caption{A $1$-$\GURA$}\label{fig:gura}%
}
\capbtabbox{%
\scalebox{.8}{
  \begin{tabular}{l||l|l}
\hline
\# registers & $\RA$ & $\URA$ \\
\hline
$*$ & undecidable~\cite{DBLP:journals/tocl/DemriL09} & {\bf in $\expspace$ (Th. \ref{thm:main})} \\
$1$ & Ackermann-cpl.~\cite{DBLP:conf/lics/FigueiraFSS11} & {\bf in $\pspace$ (Th. \ref{thm:main})} \\ 
\hline
\# registers & $\GRA$ & $\GURA$ \\
\hline 
$*$ & undecidable~\cite{KaminskiZeitlin} &  \\
$1$ & undecidable & {\bf in $\expspace$ (Th.~\ref{thm_gura}) }
\end{tabular}}
}{%
  \caption{Universality over $(\N;=)$}\label{tab:univ}%
}
\end{floatrow}
\end{figure}

In this paper, we study the \emph{universality problem}: given a $\GRA$ $\A$, is $\A$ universal, \ie, does $L(\A)=(\Sigma\times\domain)^*$ hold? 
In Table \ref{tab:univ}, we give an overview of the decidability status for register automata over $(\N;=)$, in bold the new results for unambiguous register automata that we present in this paper.

\section{Basic Notions for Deciding Universality}\label{sect:basic}
For many computational models, a standard approach for solving the universality problem is to explore the (potentially infinite) state space of the automaton under study. 
Starting from the initial state, 
the basic idea is to input one letter after the other, 
and keep track of the \emph{sets of states} that are reached, building a reachability graph whose nodes are the reached sets of states (per input letter). 
The key property of this state space is that it contains sufficient information  to decide whether the automaton under study is universal: this is the case if, and only if, every node of the graph contains an accepting state. 
Let us formalize this intuition for register automata.

Fix a $k$-$\GRA$ $\A=(\registers,\locs,\loc_{\init},\locs_\acc,\edges)$ over $\mathcal{D}$ and $\Sigma$, for some $k\in\N$.  
A \emph{configuration of $\A$} is a subset of $\locs\times\domain_\bot^k$.
The set $\config_\init$, denoting the singleton set containing the initial state of $\A$, 
is a configuration, henceforth called the \emph{initial configuration}.
Let $\config$ be a configuration, and let $(\sigma,d)\in(\Sigma\times\domain)$. 
We use $\succ_\A(\config,(\sigma,d))$ to denote the \emph{successor of $\config$ on the input $(\sigma, d)$}, formally defined by 
\begin{align*}
\succ_\A(\config,(\sigma,d)) := \{\loc(\vect{u}) \mid \exists \, \loc'(\vect{u'})\in \config \  \loc'(\vect{u'})\xrightarrow{\sigma,d}_\A\loc(\vect{u})\}.
\end{align*}
In order to extend this definition to data words, we define inductively $\succ_\A(\config,\varepsilon):=\config$ and $\succ_\A(\config, w \cdot (\sigma,d))  :=  \succ_\A(\succ_\A(\config,w),(\sigma,d))$. 
We say that a configuration $\config$ is \emph{reachable in $\A$ by the data word $w$} if $\config = \succ_\A(\config_\init, w)$; 
we say that $\config$ is \emph{reachable in $\A$} if there exists some data word $w$ such that $\config$ is reachable in $\A$ by $w$. 
We say that $\config$ is \emph{coverable} if there exists some  $\config'\supseteq \config$ such that $\config'$ is reachable in $\A$. 
Given a configuration $\config$, we use $\data(\config)$ to denote the set $\{d_i\in\domain_\bot^k\mid \exists \loc\in\locs, 1\leq i \leq k \  \loc(d_1,\dots,d_k)\in\config\}$ of data occurring in $\config$. 
Notice that every configuration reachable in an $\RA$ (without guessing) is necessarily finite. 
In contrast, %
the configuration $\{\loc_1(d') \mid d'\in\N, d'\neq d_1\}$ is reachable in the $\GRA$ in Figure~\ref{fig:gura} by the single-letter data word $(\sigma,d_1)$. 
If $\config=\{\loc(\vect{u})\}$ is a singleton set, then we may, in slight abuse of notation, omit the curly brackets and write $\loc(\vect{u})$.

We say that a configuration $\config$ is \emph{accepting} if there exists $\loc(\vect{u})\in\config$ such that $\loc\in\locs_\acc$; otherwise we say that $\config$ is \emph{non-accepting}. 
Clearly, $\A$ is universal if, and only if, every configuration reachable in $\A$ is accepting.
This suggests to reduce the universality problem to a reachability problem for the  state space corresponding to the  given input $\GRA$. 
However, the state space of a $\GRA$ is infinite, in two different aspects. 

First of all, the state space is \emph{infinitely branching}, as each of the infinite data in $\D$ may give rise to a unique successor configuration. 
The standard approach for solving this complication is to abstract from concrete data, using the simple observation that, \eg, the data word $3 \cdot 4$ is accepted from the state $\loc(4)$ if, and only if, 
$5 \cdot 2$ is accepted from the state $\loc(2)$.  
This is formalized in the following.

A \emph{partial isomorphism of $\domain_\bot$} is an injective mapping $\pi:D\to\domain_\bot$ with
domain $\textup{dom}(\pi):=D\subseteq\domain$ such that if $\bot\in D$ then $\pi(\bot)=\bot$. 
Let $\pi$ be a partial isomorphism of $\domain_\bot$ and let $\config$ be a configuration such that $\data(\config)\subseteq\textup{dom}(\pi)$.
We define the configuration $\pi(\config) := \{\loc(\pi(d_1),\dots,\pi(d_k)))\mid \loc(d_1,\dots,d_k)\in\config\}$; likewise, if $\{d_1,\dots,d_k\}\subseteq\textup{dom}(\pi)$, we define the data word  $\pi(w)=(\sigma_1,\pi(d_1))\dots(\sigma_k,\pi(d_k))$. 
We say that two pairs $\langle\config, w\rangle$ and $\langle\config',w'\rangle$ \emph{are equivalent with respect to $\pi$}, written $\langle\config,w\rangle \sim_\pi \langle\config',w'\rangle$, if
$\pi(\config)=\config' \text{ and } \pi(w)=w'$. If $w=w'=\varepsilon$, we may write $\config\sim_\pi\config'$. We write $\langle\config,w\rangle\sim\langle\config',w'\rangle$ if $\langle\config,w\rangle\sim_\pi\langle\config',w'\rangle$ for some partial isomorphism $\pi$ of $\domain_\bot$. 
\begin{proposition}
\label{prop:ra_equivalence}
Let $\A$ be a $\GRA$.
If $\langle\config,w\rangle\sim\langle\config',w'\rangle$, then
$\succ_\A(\config,w) \sim \succ_\A(\config',w')$. 
\end{proposition}
Secondly, there can be infinitely many reachable configurations even up to the equivalence relation $\sim$.
As an example, consider the $\GURA$ in Figure \ref{fig:gura}.
For every $n\geq 1$, the configuration $\config_n:=\{\loc_1(d')\mid d'\in\mN\backslash\{d_1,\dots,d_n\}\}\cup\{\loc_2(d_n)\}$ with pairwise distinct data values $d_1,\dots,d_n$ is reachable by the data word $d_1 \cdot d_2 \cdot \, \dots \, \cdot d_n$,
and $\config_n\not\sim\config_{n'}$ for $n\neq n'$.
There are similar examples also for $\URA$, cf.~\cite{DBLP:conf/stacs/MottetQ19}.

In order to obtain our results, we will prove that one can solve the reachability problem for the state space of $\A$ by focussing on a subset of configurations reachable in the automaton under study. 
The concrete methods are different for $\URA$ and $\GURA$, however, 
for both models we will take advantage of Proposition~\ref{prop:ra_equivalence} and its simple consequence (cf.~\cite{DBLP:conf/stacs/MottetQ19}). %

\begin{corollary}
\label{corollary:gnra_bad}
Let $\A$ be a $\GRA$.
If $\langle\config,w\rangle\sim\langle\config',w'\rangle$ and $\succ_\A(\config,w)$ is non-accepting (accepting, respectively), then $\succ_\A(\config',w')$ is non-accepting (accepting, respectively). 
\end{corollary}

\section{The Universality Problem for URA over $(\N;=)$}
\label{section:ura}
In this section, we study the complexity of the universality problem for $\URA$ over the relational structure $(\N;=)$. 
We prove the following theorem. 

\begin{theorem}\label{thm:main}
The universality problem is
\begin{itemize}
  \item in \pspace for $k$-$\URA$ for any fixed $k \in \N$,
  \item in \expspace for $\URA$.
\end{itemize}
\end{theorem}

We start by showing that we can assume $\URA$ to have a specific form that simplifies the coming proofs. 
Given some $k$-$\URA$ $\A$, 
we say that $\A$ is \emph{\clean} if for every state $\loc(\vect{u})$ that is reachable in $\A$ there exists a data word $w$ that is accepted from $\loc(\vect{u})$, and  $u_i\neq u_j$ for all $1\leq i<j\leq k$, \ie, no datum appears more than once in $\vect{u}$. 
The proof of the following proposition is simple and %
omitted.

\begin{proposition}
For every $k$-$\URA$ one can compute in polynomial time an equivalent \clean\  $k$-$\URA$.
\end{proposition}
In the following we always assume that a $k$-$\URA$ is \clean, even if we do not  explicitly mention it.  
For simplicity, we also assume that the alphabet of the $\URA$ we consider are singletons. 
The techniques we develop can be easily lifted to the more general case where $\Sigma$ is not a singleton.

We introduce some constants that bound from above the number of states with the same location occurring in a configuration reachable in a universal $\URA$. 
Let $\A$ be a $k$-$\URA$.  
For a configuration $\config$ of $\A$, 
define $\const_\config \in \N$ to be the maximal number $\const$ such that in $\config$ there are $\const$ different states with the
same location. 
Define $\const_\A \in \N \cup \{\infty\}$ to be the supremum of $\const_\config$, for $\config$ ranging over all the configurations $\config$ reachable in $\A$, if $\A$ is a \emph{universal} $k$-$\URA$, \ie, $L(\A)=\domain^*$.
In the sequel, we show that $\const_\A < \infty$. 
In order to do so, 
for $k \in \N$, define $\fatconst_k \in \N \cup \{\infty\}$ to be the supremum of all the $\const_\A$, for $\A$ ranging over \clean and universal $k$-$\URA$.
The main technical result of this section is showing that $\fatconst_k$ is finite and moreover upper-bounded by an exponential function of $k$.

Let $n$ be the number of locations of $\A$.  
First observe that showing $\fatconst_k \in \N$ easily implies the existence of a \npspace algorithm deciding whether $\A$ is universal.  
Indeed, if $\fatconst_k < \infty$, then every configuration $\config$ reachable in $\A$ has size at most $n \cdot \fatconst_k$,
as otherwise $\config$ contains more than $\fatconst_k$ states with the same location. Thus, in order to decide whether $\A$ is not universal, we can apply the following algorithm: 
\begin{itemize}
\item By Corollary~\ref{corollary:gnra_bad}, $\A$ is not universal iff $\A$ does not accept some data word $(\sigma_1,d_1)(\sigma_2,d_2)\dots$, where $d_i\in\{0,\dots,i\}$ for all $i$.
\item Guess, letter by letter, an input data word $(\sigma_1,d_1)(\sigma_2,d_2)\dots$, where $d_i\in\{0,\dots,i\}$.
\item For each $i\geq 1$, define $\config_i := \succ_\A(\config_{i-1},(\sigma_i,d_i))$, where $\config_0=\config_\init$. 
\item If for some $i\geq 1$, the configuration $\config_i$ is not accepting or its size exceeds $n\cdot\fatconst_k$, we know that $\A$ is not universal. 
\item Otherwise we keep the configuration in the space linear with respect to $n$ and count the length of the word.
If the length exceeds the number of possible configurations, then this run is not accepting. 
The length counter can be also kept in linear space. 
\end{itemize}
The above is hence a $\pspace$-algorithm for deciding non-universality for $k$-$\URA$. 
By Savitch's theorem, there also exists one for deciding universality for $k$-$\URA$. 
Moreover, if we show that $\fatconst_k$ is exponential in $k$, then the above algorithm works in space exponential with respect to $k$, so is in \expspace
even without fixing the number of registers $k$.
Therefore, in order to show Theorem~\ref{thm:main}, it is enough to prove that $\fatconst_k$ is bounded by some exponential function of $k$.
The rest of this Section is devoted mainly to showing the following lemma. 

\begin{lemma}\label{lem:upper}
$\fatconst_k \leq (k \cdot 4^k \cdot k!)^k$.
\end{lemma}

A short Ramsey argument given below shows that $\fatconst_k$ is finite for all $k$, however only giving a doubly-exponential bound.
Before starting the proof, 
we remark that these techniques alone \emph{cannot} be used to lower the complexity of the universality problem
for $k$-URA or for URA even more. 
This is because $\fatconst_k \geq k!$, which is the subject of the following lemma.

\begin{lemma}\label{lem:lower}
$\fatconst_k \geq k!$.
\end{lemma}
\begin{proof}
We define a family of \clean universal $k$-URA $(\A_k)_{k\geq 1}$ over $\Sigma=\{\sigma\}$ such that $\const_{\A_k} \geq k!$. 
Consider the following part of a \clean universal $k$-URA $\A_k$ (shown for the case $k=3$):

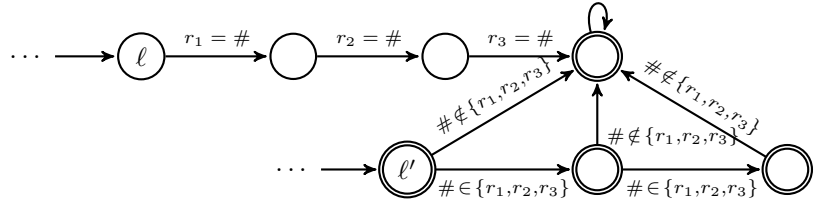
\begin{center}
\begin{tikzpicture}[->,>=stealth',shorten >=1pt,auto,node distance=4cm,thick,node/.style={circle,draw,scale=0.9}, roundnode/.style={circle, draw=black, thick, minimum size=6mm},]
\tikzset{every state/.style={minimum size=0pt}};

\node[]  (aux) at (-1.5,0) {$\dots$};
\node[roundnode]  (1) at (0,0) {$\loc$};
\node[roundnode]  (2) at (2,0) {};
\node[roundnode]  (3) at (4,0) {};
\node[roundnode,accepting]  (4) at (6,0) {};

\path [->] (aux) edge (1);
\path [->] (1) edge node[above] {\scriptsize{$\register_1=\currentinput$}} (2);

\path [->] (2) edge node[above] {\scriptsize{$\register_2=\currentinput$}} (3);

\path [->] (3) edge node[above] {\scriptsize{$\register_3=\currentinput$}} (4);

\path [->] (4) edge [loop above] (4);
\node[]  (aux1) at (2,-1.5) {$\dots$};
\node[roundnode,accepting]  (5) at (3.5,-1.5) {$\loc'$};
\node[roundnode,accepting]  (6) at (6,-1.5) {};
\node[roundnode,accepting]  (7) at (8.5,-1.5) {};

\path [->] (aux1) edge (5);
\path [->] (5) edge node[below] {\scriptsize{$\currentinput\!\in\!\{\register_1,\!\register_2,\!\register_3\}$}} (6);

\path [->] (6) edge node[below] {\scriptsize{$\currentinput\!\in\!\{\register_1,\!\register_2,\!\register_3\}$}} (7);

\path [->] (5) edge node [midway,sloped,above] {\scriptsize{$\currentinput\!\notin\!\{\register_1,\!\register_2,\!\register_3\}$}} (4);
\path [->] (6) edge node [near start,right,yshift=-1mm] {\scriptsize{$\currentinput\!\notin\!\{\register_1,\!\register_2,\!\register_3\}$}}(4);
\path [->] (7) edge node [midway,sloped,above] {\scriptsize{$\currentinput\!\notin\!\{\register_1,\!\register_2,\!\register_3\}$}}(4);
 	\end{tikzpicture}
 	\end{center}
The rest of the automaton makes sure that the configuration
\[\{\loc(\vect{u}) \mid \vect{u}\in\{1,\dots,k\}^k \text{ is a permutation}\} \cup \{\loc'(1,\dots,k)\}\] is reachable in $\A_k$ by the $k$-letter 
 data word %
$1\cdot 2 \cdot \, \dots \, \cdot k$ (\emph{e.g.}, each $\loc(\vect u)$ is reached by a path storing the input data in a different order).
By taking the (disjoint) union with an unambiguous automaton accepting every data word of length $<k$ and every $k$ letter word that has a repeated data value, we obtain a universal automaton.
\end{proof}
Our main tool to prove Lemma~\ref{lem:upper} is a structural observation, which delivers an understanding
of how reachable configurations in universal $k$-URA can look like.
Before diving into it we present an intuition by the  following example.

\begin{example}
\label{example:ura}
Let $\config$ be a configuration reachable in some universal $2$-$\URA$ $\A$ over $\Sigma=\{\sigma\}$ by some data word $w$, and assume that $\config$ 
contains three states $\loc(1,2)$, $\loc(3,4)$ and $\loc(5,6)$ sharing the same location $\loc$. 
We will argue that %
this is impossible. 
Assume that from $\loc(1,2)$ the data word $1\cdot 2\cdot 7$ is accepted. 
Then clearly
also $3\cdot 4\cdot 7$ is accepted from $\loc(3,4)$,
and $5\cdot 6\cdot 7$ is accepted from $\loc(5,6)$. 
Let us now consider the data word 
$8\cdot 9\cdot 7$,
where $8$ and $9$ are \emph{fresh} data values, that is, 
they do not occur in $w$.
Since $A$ is universal, the data word $8\cdot 9\cdot 7$ must be accepted from some state $(\loc',d_1,d_2)$ in $\config$. 
The set $\{d_1, d_2\}$ has only two elements,
and so the intersection with at least one of the sets $\{1, 2\}$, $\{3, 4\}$ and $\{5, 6\}$ must be empty.
For instance, assume that $\{d_1, d_2\} \cap \{1, 2\} = \emptyset$
and $(d_1, d_2) = (3, 6)$. 
Note that 
$\langle\loc'(3,6), 8\cdot 9\cdot 7\rangle \sim \langle\loc'(3,6), 1\cdot 2\cdot 7\rangle$, 
so that by Corollary \ref{corollary:gnra_bad}, the data word 
$1\cdot 2\cdot 7$ is accepted from state $\loc'(3,6)$, too. 
But then 
that there are two accepting runs 
for $w \cdot 1 \cdot 2 \cdot 7$, contradiction to the unambiguity of $\A$. 
Below we generalise this reasoning, in particular to the case where some registers 
in the reached states keep the same value (\ie, not all are different, as $1$, $2$, $3$, $4$, $5$ and $6$ in the above example).
However the intuition stays the same. 
\end{example}

We say that a set of tuples $T \subseteq \D_\bot^m$ is \emph{m-full} (or simply \emph{full} if $m$ is clear from the context)
if there exists a set of indices $I \subseteq [1,m]$ such that:
\begin{itemize}
  \item all the tuples in $T$ are identical in indices from $I$, namely for all $i \in I$ and all $\vect{t}, \vect{t'} \in T$ we have $t_i = t'_i$;
  \item all the data values occurring in tuples in $T$ on indices outside $I$ are different, namely for all $i\not\in I$, all $j\in\{1,\dots,m\}$, and all 
  $\vect{t}, \vect{t'} \in T$ we have $t_i \neq t'_j$ unless both $\vect{t} = \vect{t'}$ and $i = j$.
  Note that in particular, this condition applies to the case $\vect{t'}=\vect{t}$, and thus $t_i\neq t_j$ whenever $i\not\in I$ and $j\in\{1,\dots,m\}$ are different.
\end{itemize}

The following are examples of full sets:
\begin{itemize}
  \item a $4$-full set is the set of 4-tuples $(1, 2, 3, 4), (1, 2, 5, 6), (1, 2, 7, 8)$, in that case $I = \{1, 2\}$;
  \item a $5$-full set containing one tuple $(3, 7, 2, 10, 8)$, any set of indices $I \subseteq [1,5]$ works here;
  \item a $2$-full set containing tuples $(2, 1), (3, 1), (4, 1), (5, 1)$, in that case $I = \{2\}$.
\end{itemize}

For a location $\loc$ and set of tuples $T \subseteq \D_\bot^k$ we write $\loc(T) = \{\loc(\vect{t}) \mid \vect{t} \in T\}$.
The following lemma delivers the key observation, which uses the notion of $k$-full sets.

\begin{lemma}\label{lem:structure}
If $\A$ is a \clean universal $k$-$\URA$, 
then there exists no configuration $\config$  reachable in $\A$ such that %
 $\loc(T) \subseteq \config$
for some location $\loc\in\locs$ 
and some k-full set of tuples $T \subseteq \D_\bot^k$ of size more than $k$.  
\end{lemma}

\begin{proof}
Let $\A$ be  a \clean universal $k$-$\URA$, and 
suppose towards contradiction 
that there exists a configuration $\config$ reachable in $\A$ such that 
$\loc(T) \subseteq \config$ for some location $\loc$ and some $k$-full set $T \subseteq \D_\bot^k$ of size more than $k$. 
Let $w$ be the data word such that $\config=\succ_\A(\config_\init,w)$. 
Assume without loss of generality that the indices on which tuples from $T$ are identical are $I = \{1, \ldots, n\}$ for some $n \leq k$.
Let us choose some $k+1$ tuples from $T$, let the $i$-th of it be of the form $\vect{t}^i = (c_1, \ldots, c_n, o_1^i, \ldots, o_m^i)$, where
$n+m = k$. 
We call the $c_j$ the \emph{common} data values and the $o_j^i$ the \emph{own} data values of $\vect{t}^i$. 
$\A$ is \clean \ and $\loc(\vect{t}^1)$ is reachable in $\A$, so there must  exist a data word $w_1 \in (\Sigma\times\D)^*$ that 
is accepted from $\loc(\vect{t}^1)$. 
Without loss of generality we can assume that $w_1$ does not contain the own data values of any of the other tuples $\vect{t}^2,\dots, \vect{t}^{k+1}$. Indeed, if this is the case, we can replace synchronously all occurrences of such a data value by a \emph{fresh} data value not occurring in $\data(w)$; the resulting data word is still accepted from $\loc(\vect{t}^1)$. 
For every $i \in [2,k+1]$, let $w_i$ be the word $w_1$ in which for each $j \in [1,m]$ the own data
value $o_j^1$ is replaced by the data value $o_j^i$. 
Clearly, for every $i \in [2,k+1]$
$\langle\loc(\vect{t}^1),w_1\rangle \sim \langle \loc(\vect{t}^i),w_i\rangle$, 
so that by Corollary \ref{corollary:gnra_bad} 
the data word $w_i$ is accepted from $\loc(\vect{t}^i)$.

Let us now consider the data word $w_\fresh$ that is obtained
from $w_1$ by replacing synchronously every occurrence of 
every $o_j^1$ is by some fresh data value each. 
As $\A$ is universal,  also the data word $w_\fresh$ needs to be accepted from  some state in $\config$.
Let $q_\fresh = \loc'(e_1, \ldots, e_k)$ be the state in $\config$ from which $w_\fresh$ is accepted.
Notice that we do not enforce $\loc \neq \loc'$, similarly $e_i$ may be equal to some of the $c_i$
or $o_i^j$, but this does not have an effect on our reasoning. 
For each tuple $\vect{t}^i = (c_1, \ldots, c_n, o_1^i, \ldots, o_m^i)$,
let the
set of its own data values be $O_i = \{o_1^i, \ldots, o_m^i\}$. 
By assumption all the sets $O_1, \ldots, O_{k+1}$ are pairwise
disjoint. 
As there are $k+1$ of them, we know that at least one of them is disjoint from the set of data values in the state
$q_\fresh$, namely with $E = \{e_1, \ldots, e_k\}$. Assume without loss of generality that $O_1 \cap E = \emptyset$. 
This however means that 
$\langle q_\fresh,w_1\rangle \sim \langle q_\fresh,w_\fresh\rangle$, 
so that by Corollary \ref{corollary:gnra_bad} $w_1$ 
is also accepted from $q_\fresh$. 
In consequence, there are at least two accepting runs over $w_1$ from configuration $\config$,
one from $\loc(\vect{t}^1)$ and one from $q_\fresh$. 
Hence there are at least two initialized accepting runs over $w\cdot w_1$.
This is a contradiction to the unambiguity of $\A$. 
\end{proof}

We give here the argument showing that $\fatconst_k$ is bounded by some doubly-exponential function in $k$.
We show this argument in order to illustrate the techniques, which needed to be refined in our proof of Lemma~\ref{lem:upper}.
First recall that the \emph{Ramsey number} $R_m(n)$ is the smallest number of vertices $k$ of the graphs such that 
any clique of $k$ vertices with its edges coloured on $m$ different colours contain a monochromatic subgraph $G$ of $n$ vertices,
namely such that all the edges in $G$ are of the same colour.
It can be shown by induction that $R_m(n)$ is finite, and indeed its growth is bounded by $2^{n^{O(m)}}$.
For the definition of \emph{$k$-full} set consider page 8.

\begin{restatable}{proposition}{ramsey}\label{prop:ramsey}
Every set $T \subseteq \D_\bot^k$ of size at least $R_{k+1}(4^k (k+1)!+1)$ contains a k-full subset of size at least $k+1$.
\end{restatable}
\begin{proof}
Construct a graph with vertices being tuples from $T$ and edge between $\vect{t}$ and $\vect{t'}$ be coloured by the number of data values that $\vect{t}$ and $\vect{t'}$ have in common.
Clearly the colour belongs to the set $\{0, \ldots, k\}$, so there are $k+1$ colours.
Because $|S| \geq R_{k+1}(4^k (k+1)!+1)$ we know from Ramsey's theorem that there are at least $4^k (k+1)!+1$ tuples such that every
intersection is of the same size - assume this size to be $m$.
Let $S$ be a set of $4^k (k+1)!+1$ such tuples and let $\vect{s} \in S$ be one of them. Let $\vect{s} = (d_1, \ldots, d_k)$. Divide all the other tuples $\vect{s'}$
into $\binom{k}{m}^2 \cdot m!$ sets depending on which $m$ data values from $\{d_1, \ldots, d_k\}$ belong to $\vect{s'}$
(there are $\binom{k}{m}$ options), on which positions they are located in $\vect{s'}$ (also $\binom{k}{m}$ options)
and in which order ($m!$ options). It is easy to see that $\binom{k}{m}^2 \cdot m! \leq 4^k k!$, as $\binom{k}{m} \leq 2^k$ and $m! \leq k!$.
We divide $4^k (k+1)!$ tuples (we omit $\vect{s}$) into at most $4^k k!$ sets,
so by the pigeonhole principle at least one of them contains at least $k+1$ elements: let these elements be $\vect{s^1}, \ldots,\vect{s^{k+1}}$.
Notice now that the tuples $\vect{s^1}, \ldots, \vect{s^{k+1}}$ form a $k$-full set:
indeed on positions on which they have the $m$ shared data they are identical
and on the other positions all the data values are totally different.
Thus $T$ contains a $k$-full set of size $k+1$, which finishes the proof.
\end{proof}

By refining the reasoning, we obtain the following result that directly implies Lemma~\ref{lem:upper}, when setting $B=n=k$.

\begin{restatable}{lemma}{twoparameters}\label{lem:two-parameters}
Every set $T \subseteq \D_\bot^n$ of size at least $(B\cdot 4^n \cdot n!)^n + 1$ contains an $n$-full subset of size bigger than $B$.
\end{restatable}
\begin{proof}
Let us denote by $D_{B, n}$ the maximal size of the set of $n$-tuples such that any $n$-full subset has size at most $B$; 
in other words, $D_{B,n}$ is the least integer such that if $X$ is a set of $n$-tuples of size $D_{B,n}+1$, then $X$ contains an $n$-full subset of size $B+1$.
Our aim is to show that $D_{B,n} \leq (B \cdot 4^n \cdot n!)^n$.  We show it by induction on $n$.

For the induction base assume $n = 1$. Then any set of data values is a full set, so clearly $D_{B, 1} \leq B \leq B\cdot 4^1 \cdot 1!$.

Assume now that $D_{B, m} \leq (B \cdot 4^m \cdot m!)^m$ for all $m < n$ and consider some set $T \subseteq \D_\bot^n$ of $n$-tuples.
Assume that $T$ contains no $n$-full subset of size bigger than $B$. 
Pick some tuple $\vect{t} = (d_1, \ldots, d_n) \in T$.
We first show that there can be at most $4^n \cdot n! \cdot D_{B,n-1}$ tuples in $T$ whose data intersect $\ds(\vect{t})$.
Let us denote $N = 4^n \cdot n! \cdot D_{B,n-1}$.
Let $S$ be the set of those tuples, assume towards contradiction that the size of $S$ exceeds the bound $N$. 
For each tuple $\vect{s} \in S$ there are at most $2^n-1$ choices for $\ds(\vect{s}) \cap \ds(\vect{t})$,
so by the pigeonhole principle there are more than $2^n \cdot n! \cdot D_{B,n-1}$ tuples which have the same set $\ds(\vect{s}) \cap \ds(\vect{t})$.
Those data values can occur in tuples from $S$ on at most $2^n$ different sets of indices, and in at most $n!$ different orders,
so by the pigeonhole principle more than $D_{B,n-1}$ tuples from $S$ have the same data values shared with $\vect{t}$ on the same indices.
After ignoring the indices shared with $\vect{t}$ at most $n-1$ indices remain on these tuples. So by induction assumption there is
some full set of size more than $B$ among these tuples, which leads to the contradiction with assumption that for more than $N$
tuples from $T$ their data intersects $\ds(\vect{t})$.

Therefore we know that all the tuples but the mentioned $N$ ones have data  disjoint with $\ds(\vect{t})$.
Let use denote $\vect{t^1} = \vect{t}$ and $T_1$ to be the set of tuples with data disjoint from $\ds(\vect{t^1})$.
Let $\vect{t^2} \in T_1$.
We now repeat the argument for $\vect{t^2}$ similarly as for $\vect{t^1}$ and get that there are at most $N$
tuples with data intersecting $\ds(\vect{t^2})$. Repeating this argument we get a sequence of tuples $\vect{t^1}, \vect{t^2}, \ldots, \vect{t^m}$ such
that for each $i \neq j$ we have $\ds(\vect{t^i}) \cap \ds(\vect{t^j}) = \emptyset$. After adding each tuple $\vect{t^j}$ to the sequence we define
the set $T_{j+1}$ of elements, which have disjoint data with all the tuples $\vect{t^1}, \ldots, \vect{t^j}$. As long as $T_{j+1}$ is nonempty we
can continue the process. It is easy to see that $|T_{j+1}| \geq |T_j| - N$. Assume now towards contradiction that
$D_{B,n} > (B \cdot 4^n \cdot n!)^n$, which implies that $D_{B,n} > (B \cdot 4^n \cdot n!) \cdot D_{B,n-1} = B \cdot N$.
We can see now that $|T_B| > 0$, which means that we can construct tuples $\vect{t^1}, \vect{t^2}, \ldots, \vect{t^B}, \vect{t^{B+1}}$
such that for each $i \neq j$ we have $\ds(\vect{t^i}) \cap \ds(\vect{t^j}) = \emptyset$. This however means that $\{\vect{t^1}, \ldots, \vect{t^{B+1}}\}$
is a full set of size $B+1$, which is more than $B$. This contradicts the assumption, which shows that $D_{B,n} \leq (B \cdot 4^n \cdot n!)^n$ and finishes the proof.
\end{proof}

We can now apply a reduction from containment to universality provided by Barloy and Clemente (Lemma 8 in~\cite{BarloyClemente}) to obtain Theorem~\ref{thm:urabound} from the introduction.
\section{Universality for URA over $(\Q;<,=)$} 
In this section, we prove Theorem~\ref{thm:uraorderbound} by using the techniques developed in the preceding section. 
Let us define constants $\fatconst^\Oo_k$ for $k$-URA with order similarly as $\fatconst_k$ for $k$-URA. The main technical lemma is the following; Theorem~\ref{thm:uraorderbound} follows. 
\begin{lemma}
$\fatconst^\Oo_1 = 1$.
\end{lemma}
\begin{proof}
Towards contradiction suppose that for some \clean universal 1-URA  $\A$  with order there is a configuration $\config$ reachable in $\A$ 
by a data word $w_\pref \in (\Sigma\times\Q)^*$, 
such that $\loc(d_1), \loc(d_2) \in \config$ for some location $\loc$ and data values $d_1 < d_2$.
Because $\A$ is \clean,
there exists a data word $w_1 \in (\Sigma\times\Q)^*$ that is accepted from $\loc(d_1)$. 
Without loss of generality we can assume that $w_1$ does not contain any data in $(d_1,d_2]$. 
Indeed, if $w_1$ contains some datum in $(d_1,d_2]$, 
then we can replace it synchronously by some datum greater than $d_2$, while taking care that that the relative order of all data in $w_1$ is preserved, so that, for the resulting data word $w$, we have $\langle \loc(d_1),w_1 \rangle \sim \langle \loc(d_1),w\rangle$. 
By Corollary \ref{corollary:gnra_bad}, the resulting data word $w$ is also accepted from $\loc(d_1)$. 
Notice that for similar reasons, also the data word $w_2$ obtained from $w_1$ by replacing every occurrence of $d_1$ by $d_2$ is accepted from $\loc(d_2)$. 
Now, if $w_1$ does not contain $d_1$, then $w_1 = w_2$. 
Hence $w_1$ is accepted from both $\loc(d_1)$ and $\loc(d_2)$, contradiction to unambiguity of $\A$. 
So let us assume $w_1$ contains $d_1$. 
Pick some data value $d_\fresh$ that is fresh, \ie, it does not occur in $w_\pref$, and additionally $d_1 < d_\fresh < d_2$. 
We clearly can choose such a fresh data value, as there are infinitely many rational numbers between $d_1$ and $d_2$
and only finitely many of them occur in $w_\pref$. Let $w_\fresh$ be the word obtained from $w_1$ by synchronously replacing every occurrence of $d_1$ by  $d_\fresh$.
The word $w_\fresh$ is accepted from some configuration in $\config$, let it be $\loc'(d')$. %
Notice now that if $d' < d_\fresh$,
then $\langle \loc'(d'),w_\fresh \rangle \sim \langle \loc'(d'),w_2\rangle$, so that 
 $\loc'(d')$ accepts also $w_2$ by Corollary \ref{corollary:gnra_bad}; in the other case, \ie, if $d' > d_\fresh$, then 
 we have $\langle\loc'(d'),w_\fresh \rangle \sim \langle\loc'(d'),w_1\rangle$, so that $\loc'(d')$ also 
accepts $w_1$. Therefore in the first case automaton $\A$ has two accepting runs over $w_\pref \cdot w_2$ and in the second case over $w_\pref \cdot  w_1$.
This is a contradiction to the unambiguity of $\A$.
\end{proof}

The following lemma shows that our techniques by itself are not sufficient to solve the case of 2-URA with order.

\begin{restatable}{lemma}{orderedinfinity}
$\fatconst^\Oo_2 = \infty$.
\end{restatable}
\begin{proof}
For all $n\geq 1$, consider the configuration $C_n:=\{\loc'(1,n),\loc(1,2),\dots,\loc(n-1,n)\}$, which is for all $n\geq 1$ a subset of a configuration reachable in the following \clean universal 2-URA.
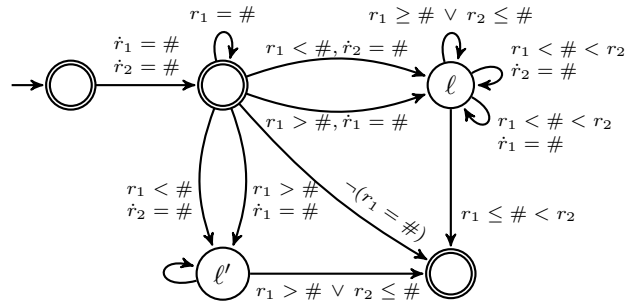
\begin{center}
\begin{tikzpicture}[->,>=stealth',shorten >=1pt,auto,node distance=4cm,thick,node/.style={circle,draw,scale=0.9}, roundnode/.style={circle, draw=black, thick, minimum size=6mm},]
\tikzset{every state/.style={minimum size=0pt}};

\node[roundnode, accepting, ,initial, initial text={}]  (1) at (0,0) {};
\node[roundnode,accepting]  (2) at (2,0) {};
\node[roundnode]  (3) at (5,0) {$\loc$};
\node[roundnode]  (4) at (2,-2.5) {$\loc'$};
\node[roundnode, accepting]  (5) at (5,-2.5) {};

\path [->] (1) edge node[above] {\scriptsize{$\dot\register_2=\currentinput$}} 
node [above,yshift=3mm] {\scriptsize{$\dot\register_1=\currentinput$}} (2);

\path [->] (2) edge [loop above] node[above] {\scriptsize{$\register_1=\currentinput$}} (2);

\path [->] (2) edge[bend left=20] node[above,yshift=-1mm] {\scriptsize{$\register_1<\currentinput, \dot\register_2=\currentinput$}} (3);

\path [->] (2) edge[bend right=20] node[below,yshift=1mm] {\scriptsize{$\register_1>\currentinput, \dot\register_1=\currentinput$}} (3);

\path [->] (2) edge[bend right=20,near end] node[left,yshift=3mm] {\scriptsize{$\register_1<\currentinput$}} node[left] {\scriptsize{$\dot\register_2=\currentinput$}} (4);

\path [->] (2) edge[bend left=20,near end] node[right,yshift=3mm] {\scriptsize{$\register_1>\currentinput$}} node[right] {\scriptsize{$\dot\register_1=\currentinput$}} (4);

\path [->] (4) edge node[below] {\scriptsize{$\register_1>\currentinput \, \vee \, \register_2\leq \currentinput$}} (5);

\path [->] (3) edge node[right,near end] {\scriptsize{$\register_1\leq \currentinput <\register_2$}} (5);

\path [->] (4) edge [loop left] (4);

\path [->] (3) edge [loop above]  node[above] {\scriptsize{$\register_1\geq \currentinput \, \vee \, \register_2\leq \currentinput$}} (3);

\path [->] (3) edge[out=390,in=360,looseness=8]  node[right,yshift=3mm] {\scriptsize{$\register_1< \currentinput < \register_2$}} node[right] {\scriptsize{$\dot\register_2=\currentinput$}} (3);

\path [->] (3) edge[out=330,in=300,looseness=8] node[right] {\scriptsize{$\register_1< \currentinput < \register_2$}} node[right,yshift=-3mm] {\scriptsize{$\dot\register_1=\currentinput$}} (3);

\path [->] (2) edge[bend right=9] node[above,near end,sloped] {\scriptsize{$\neg(\register_1=\currentinput) $}} (5);
 	\end{tikzpicture}
 	\end{center}
The state $\loc'(d_1,d_2)$ keeps track of the first two distinct data read, with $d_1<d_2$. It is responsible for accepting any datum $d$ outside the interval $[d_1,d_2)$.
The state $\loc(x,y)$ is such that $d_1\leq x<y\leq d_2$ and it is responsible for accepting every datum $d'$ in the interval $[x,y)$.
Moreover, if $d'\in(x,y)$, then $\loc(x,y)$ splits into $\loc(x,d')$ and $\loc(d',y)$.
The automaton ensures that no two intervals $[x,y),[x',y')$ overlap, and that all the intervals $[x,y)$ present in a configuration cover the interval $[d_1,d_2)$, thus the automaton is unambiguous and universal.
\end{proof}

\section{Containment for GURA over $(\N;=)$}
In this section, 
we aim to prove the decidability of the universality problem for the more expressive model of $\GURA$. 
Let us first argue that the techniques developed in Section \ref{section:ura} do not work for $\GURA$.
\begin{example}
One can easily construct a universal $1$-$\GURA$ with reachable configuration $\config$ containing $\loc(0), \loc(1)$, and $\loc' \times \{n\in\N \mid n\neq 0,1\}$. 
If from both $\loc$ and $\loc'$ there are outgoing edges with constraint $\register=\currentinput$ to some accepting state, then every data word $n$ is accepted from $\config$. In particular, the word $0$ is accepted from $\loc(0)$, but we cannot replace $0$ by some \emph{fresh} datum to obtain a contradiction  as in Example \ref{example:ura}. 
\end{example}
The example shows that we need more sophisticated methods to solve the universality problem.
Moreover, and in contrast to the result for $\RA$, %
we cannot rely on the reduction from containment to universality by Barloy and Clemente~\cite{BarloyClemente}, as it holds for $\RA$ without guessing only. %
We hence present a direct proof for containment as stated in Theorem  \ref{thm_gura}. 
The idea is based on exploring a sufficiently big part of the infinite  \emph{synchronized state space} of both automata $\A$ and $\B$, following the approach in~\cite{DBLP:conf/stacs/MottetQ19}. 
The main difference with~\cite{DBLP:conf/stacs/MottetQ19} lies in the complications that arise due to the fact that a configuration of a $\GURA$ may be \emph{infinite}.

\subsection{Synchronized Configurations and Bounded Supports}
For the rest of this section, let  $\A=(\registers^\A,\locs^\A,\loc^\A_{\init},\locs^\A_{\acc},\edges^\A)$ be a \GRA\ with $\registers^A=\{\register_1,\dots,\register_m\}$, and let 
$\B=(\registers^\B,\locs^\B,\loc^\B_{\init},\locs^\B_{\acc},\edges^\B)$ be a \GURA\  with a single register $\register$.

We aim to reduce the containment problem $L(\A)\subseteq L(\B)$ to a reachability problem in $(\sNodes,\sTo)$ where:
\begin{itemize}
\item $\sNodes$ is the set of \emph{synchronized configurations} $(\loc(\vect{d}),\config)$, where $\loc(\vect{d})\in (\locs^\A\times\N^{\registers^\A}_\bot)$ is a single state of $\A$, and
$\config$ is a configuration of $\B$,
\item $(\loc(\vect{d}),\config)\sTo (\loc'(\vect{d'}),\config')$ if there exists a letter $(\sigma,d)\in(\Sigma\times\N)$ such that $\loc(\vect{d})\xrightarrow{\sigma,d}_\A \loc'(\vect{d'})$, and 
$\succ_\B(\config,(\sigma,d))=\config'$.
\end{itemize}
We define $\sconfig_{\init}:=(\loc_{\init}^\A(\vect{v}_\init), \config_\init)$ to be the \emph{initial synchronized configuration of $\A$ and $\B$}.
We say that a synchronized configuration $\sconfig'$ is \emph{reachable} from $\sconfig$ if there is a $\sTo$-path from $\sconfig$ to $\sconfig'$. $\sconfig$ is \emph{reachable} if it is reachable from $\sconfig_\init$.
Call a synchronized configuration $(\loc(\vect{d}),\config)$ \emph{bad} if $\loc\in\locs_\acc^\A$ is an accepting location and $\config$ is non-accepting, \ie, $\loc'\not\in\locs_\acc^\B$ for all $(\loc',u)\in\config$. 
Thus, a bad synchronized configuration is reachable iff $L(\A)\not\subseteq L(\B)$.

We extend the equivalence relation $\sim$ defined in Section~\ref{sect:basic} to synchronized configurations in a natural manner, \ie,  given a partial isomorphism $\pi$ of $\N_\bot$
such that $\data(\vect{d})\cup\data(\config)\subseteq\textup{dom}(\pi)$,
we define $(\loc(\vect{d}),\config) \sim_\pi (\loc(\vect{d}'),\config')$ if $\pi(\config)=\config'$ and $\pi(\vect{d})=\vect{d}'$.
We shortly write $\sconfig\sim\sconfig'$ if there exists a partial isomorphism $\pi$ of $\N_\bot$ such that $\sconfig\sim_\pi\sconfig'$.
Clearly, an analogon of Corollary~\ref{corollary:gnra_bad} holds for this extended relation. In particular, we have the following:
\begin{proposition}\label{prop:equivalence-relation-synch-compatible}
Let $\sconfig,\sconfig'$ be two synchronized configurations of $(\sNodes,\sTo)$ such that $\sconfig\sim\sconfig'$.
If $\sconfig$ reaches a bad synchronized configuration, so does $\sconfig'$.
\end{proposition}

The \emph{support} of a configuration $\config$ of $\B$ is the set $\supp(\config)$ of data $d'$ such that at least one of the following two conditions holds:
\begin{itemize}
\item $\loc(d')\in\config$ for some $\loc\in\locs$ such that $(\{\loc\}\times\domain)\cap\config$ is finite, 
\item $\loc(d')\not\in\config$ for some $\loc\in\locs$ such that $(\{\loc\}\times\domain)\cap\config$ is cofinite. 
\end{itemize}
Note that $\supp(\config)\subseteq\data(w)$ whenever $\config = \succ_\A(\config_\init,w)$.%

Let $\sconfig=(\loc(\vect{d}),\config)$ be a synchronized configuration, and let $a,b\in\supp(\config)$ be two data values in the support of $\config$. We say that \emph{$a$ and $b$ are indistinguishable in $\sconfig$}, written $\indiscernible{a}{b}{\sconfig}$, 
if $a,b\not\in\data(\vect{d})$ and $\{\loc \in\locs \mid \loc(a)\in C\} = \{\loc\in\locs\mid \loc(b)\in C\}$.

Given a configuration $\config$ of $\B$, we define for every datum $d\in\N$ the sets
\begin{align*}
\config_d^+ &:=  \, \{\loc(d)\in \locs\times\{d\} \mid \loc(d)\in\config \text{ and } \data(\config\cap (\{\loc\}\times\N)) \text{ is finite}\}\\
\config_d^- &:=  \, \{\loc(d)\in \locs\times\{d\} \mid \loc(d)\not\in\config \text{ and } \data(\config\cap (\{\loc\}\times\N)) \text{ is infinite}\}.
\end{align*}

We give here an example for the definition of $\config^+_d$ and $\config^-_d$.
\begin{example}
Let $\config = \{\loc_1(0),\loc_1(1)\} \cup \{\loc_2(d) \mid d\in\mN\backslash\{1,2\}\}\cup\{\loc_3(d)\mid d\in\mN\backslash\{0,1\}\}$. Then
\begin{center}
\begin{tabular}{lll}
$\config^+_0 = \{\loc_1(0)\}$ & $\config^+_1 = \{\loc_1(1)\}$ & $\config^+_2=\emptyset$ \\
$\config^-_0=\{\loc_3(0)\}$ & $\config^-_1 = \{\loc_2(1),\loc_3(1)\}$ & $\config^-_2=\{\loc_2(2)\}$
\end{tabular}
\end{center}
\end{example}
We say that a configuration $\config$ is \emph{essentially coverable} if for every two $\loc(u),\loc'(u')\in\config$, the set $\{\loc(u),\loc'(u')\}$ is coverable. 
\begin{restatable}{proposition}{esscov}\label{prop:esscov}
Let $\config$ be an essentially coverable configuration, and let $b\in\supp(\config)$.
Then $((\config\backslash\config^+_b)\cup\config^-_b)$ is essentially coverable, too.
\end{restatable}
\begin{proof}
Let $\loc(c),\loc'(c')\in ((\config\backslash\config^+_b)\cup\config^-_b)$. 
If $\loc(c),\loc'(c')\in\config\backslash\config^+_b$, then
$\{\loc(c),\loc'(c')\}$ is coverable by essential coverability of $\config$. 
Suppose $\loc(c),\loc'(c')\in\config^-_b$. By definition of $\config^-_b$, $c=c'=b$. 
Pick some value $e\in\N\backslash\{b\}$ such that $\loc(e),\loc'(e)\in\config$. Note that such a value $e$ must exist, as by definition of $\config^-_b$, the sets $\data((\{\loc\}\times\N)\cap C)$ and $\data((\{\loc'\}\times\N)\cap C)$ are cofinite, and hence their intersection is non-empty.
By essential coverability of $\config$, $\{\loc(e),\loc'(e)\}$ is coverable. 
There must thus exist some data word $w$ such that $\{\loc(e),\loc'(e)\}\subseteq \succ(\loc_\init(\bot),w)$. 
Let $\pi$ be any partial isomorphism satisfying $\pi(e)=b$ and whose domain contains $\data(w)$. 
Clearly, $\{\loc(b),\loc'(b)\} \subseteq \succ(\loc_\init(\bot),\pi(w))$, and hence $\{\loc(b),\loc'(b)\}$ is coverable. 
Finally, suppose $\loc(c)\in\config\setminus C_b^+$ and $\loc'(c')\in\config^-_b$. By definition, we have $c\neq b=c'$.
Since $\data(({\loc'}\times\N)\cap C)$ is cofinite, there is $d\neq c$ such that $\loc'(d)\in C$.
By essential coverability of $C$, there exists a data word $w$ such that $\{\loc(c),\loc'(d)\}\subseteq \succ(\loc_\init(\bot),w)$.
By picking a partial isomorphism $\pi$ such that $\pi(d)=c'$ and $\pi(c)=c$, we obtain that $\{\loc(c),\loc'(c')\}\subseteq \succ(\loc_\init(\bot),\pi(w))$, which concludes the proof.
\end{proof}

The following is the main technical result of this section. 
\begin{restatable}{proposition}{collapsegura}\label{prop:collapse_gura}
Let $\sconfig=(\loc^\A(\vect{d}),\config)$ be a synchronized configuration of $\A$ and $\B$ such that $\config$ is essentially coverable, and let $a\neq b$ be such that $a,b\in\supp(\config)$ and $\indiscernible{a}{b}{\sconfig}$.
Then $\sconfig$ reaches a bad configuration in $(\sNodes,\sTo)$ if, and only if, $S':=(\loc^\A(\vect{d}),(\config\setminus \config_b^+)\cup \config_b^-)$ reaches a bad configuration in $(\sNodes,\sTo)$.
\end{restatable}
\begin{proof}
$(\Leftarrow)$
	Suppose there exists some data word $w$ such that
	 there exists an accepting run of $\A$ on $w$ that starts in $\loc^\A(\vect{d})$,  and $\succ_\B(\config\backslash \config_b^+\cup \config_b^-,w)$ is non-accepting. 
	We assume in the following that $\succ_\B(C^+_b,w)$ is accepting; otherwise we are done. 
	Let $\loc^+(b)\in C_b^+$ be the unique state such that $\succ_\B(\loc^+(b),w)$ is accepting. 	
	In the following, we prove that we can without loss of generality assume that $w$ does not contain any $a$'s. 
	Pick some $a'\in\N$ such that $a'\not\in\data(w)\cup\supp(\config)\cup\data(\vect{d})$. 
	Let $\pi$ be the isomorphism defined by $\pi(a)=a'$, $\pi(a')=a$, and $\pi(d)=d$ for all $d\in\N_\bot\backslash\{a,a'\}$.  
	Then 
	$\langle\loc^\A(\vect{d}),w\rangle \sim_\pi\langle\loc^\A(\vect{d}),\pi(w)\rangle$ (as $a\not\in\data(\vect{d})$ by $\indiscernible{a}{b}{\sconfig}$), and 
	$\langle\loc^+(b),w\rangle \sim_\pi \langle\loc^+(b),\pi(w)\rangle$.  
	By Corollary \ref{corollary:gnra_bad}, there exists an accepting run of $\A$ on $\pi(w)$ that starts in $\loc^\A(\vect{d})$, and  $\succ_\B(\loc^+(b),\pi(w))$ is accepting. 
	We prove that $\succ_\B(\loc(c),\pi(w))$ is non-accepting, for every $\loc(c)\in\config\setminus\{\loc^+(b)\}\cup\config^-_b$:  
	first, let $\loc(c)\in C\setminus\{\loc^+(b)\}$. 
	By essential coverability of $\config$, $\{\loc^+(b),\loc(c)\}$ is coverable. 
	By unambiguity of $\B$, $\succ_\B(\loc(c),\pi(w))$ must be non-accepting. 
	Second, let $\loc(c)\in\config^-_b$. But then $c=b$, and hence $\langle\loc(c),w\rangle \sim_\pi \langle\loc(c),\pi(w)\rangle$. By assumption, $\succ_\B(\loc(c),w)$ is non-accepting, so that by Corollary \ref{corollary:gnra_bad},  $\succ_\B(\loc(c),\pi(w))$ is non-accepting, too. 
	Note that $\pi(w)$ indeed does not contain any $a$'s. 
	We can hence continue the proof assuming that 
	$w$ does not contain any $a$'s. 
	
	Next, we prove that if we replace all $b$'s occurring in $w$ by some fresh datum not occurring in $\supp(\config)\cup\data(w)\cup\data(\vect{d})$, we 
	obtain a data word that guides $\sconfig$ to a bad synchronized configuration. 
	Formally, pick some datum $b'\not\in\data(w)\cup\supp(\config)\cup\data(\vect{d})$, and let $\pi$ be the isomorphism defined by 
	$\pi(b)=b'$, $\pi(b')=b$, and $\pi(d)=d$ for all $d\in\N_\bot\backslash\{b,b'\}$. 
	Note that $\pi(w)$ does not contain any  $a$'s or $b$'s. 
	Clearly, 
	$\langle\loc^\A(\vect{d}),w\rangle \sim_\pi\langle\loc^\A(\vect{d}),\pi(w)\rangle$. By Corollary \ref{corollary:gnra_bad}, there still exists an accepting run of $\A$ on $\pi(w)$ that starts in $\loc^\A(\vect{d})$. 	
	We prove that $\succ_\B(C,\pi(w))$ is non-accepting. 
	Let $\loc(c)\in C$. 
	We distinguish three cases. 
	\begin{enumerate}
	\item Let $c\not\in\{b,b'\}$. 
	Then $\langle\loc(c),w\rangle\sim_\pi \langle\loc(c),\pi(w)\rangle$. 
	Since $\succ_\B(\loc(c),w)$ is non-accepting by assumption, so that by Corollary \ref{corollary:gnra_bad} also $\succ_\B(\loc(c),\pi(w))$ is non-accepting. 
	
	\item Let $c=b$. By $\indiscernible{a}{b}{\config}$, the state $\loc(a)$ is in $C$ and $\loc(a),\pi(w)\sim \loc(c),\pi(w)$
	since $a$ and $c$ do not appear in $w$. By essential coverability of $\config$, $\{\loc(a),\loc(c)\}\subseteq\config$ is coverable. 
	By unambiguity of $\B$, we obtain that $\succ_\B(\loc(c),\pi(w))$ is non-accepting.

	\item Let $c=b'$. Note that $\langle\loc(b),w\rangle \sim_\pi \langle\loc(b'),\pi(w)\rangle$.  
	Recall that $b'\not\in\supp(\config)$.
	This implies that $\data(\config\cap(\{\loc\}\times\N_\bot))$ is cofinite. 
	We distinguish two cases. 
	\begin{itemize}
	\item $b\in\data(\config\cap(\{\loc\}\times\N_\bot))$, i.e., $\loc(b)\in\config$. But note that $\loc(b)\not\in\config^+_b$ by cofiniteness of $\data(\config\cap(\{\loc\}\times\N_\bot))$.
	Hence $\loc(b)\in \config\backslash\{\loc^+(b)\}$.  
	\item  $b\not\in\data(\config\cap(\{\loc\}\times\N_\bot))$, i.e., $\loc(b)\in\config^-_b$. 
	\end{itemize}
	In both cases, we have proved above that $\succ(\loc(b),w)$ is non-accepting. By  $\langle\loc(b),w\rangle \sim_\pi \langle\loc(b'),\pi(w)\rangle$ and Corollary \ref{corollary:gnra_bad}, $\succ_\B(\loc(b'),\pi(w))$ is non-accepting, too.
	\end{enumerate}
	Altogether we have proved that $\succ_\B(\config,\pi(w))$ is non-accepting, while there exists some accepting run of $\A$ on $\pi(w)$  starting in $\loc^\A(\vect{d})$. 
	This concludes the proof for the $(\Leftarrow)$-direction.
	
	$(\Rightarrow)$ 
	Suppose there exists some data word $w$ such that  there exists some accepting run of $\A$ on $w$ starting in $\loc^\A(\vect{d})$, and $\succ_\B(\config,w)$ is non-accepting. 
	We assume in the following that $\succ_\B(\config\setminus \config_b^+\cup \config_b^-,w)$ is accepting; otherwise we are done. 
	Let $\loc^-(b)$ be a state in $\config_b^-$ such that $\succ_\B(\loc^-(b),w)$ is accepting. 
	Pick some datum $a'\in\N_\bot$ such that $a'\not\in\data(w) \cup \supp(\config)\cup\data(\vect{d})$. 
	Let $\pi$ be the isomorphism defined by $\pi(b)=a$, $\pi(a)=a'$, $\pi(a')=b$, and $\pi(d)=d$ for all $d\in\N\backslash\{a,b,a'\}$. 
	Clearly, $\langle\loc^\A(\vect{d}),w\rangle \sim_\pi\langle\loc^\A(\vect{d}),\pi(w)\rangle$, so that by Corollary \ref{corollary:gnra_bad},  there exists some accepting run of $\A$ on $\pi(w)$ starting in $\loc^\A(\vect{d})$. 
	We prove that $\succ_\B(\config\backslash\config^+_b\cup\config^-_b,\pi(w))$ is non-accepting. 
	Let $\loc(c)\in \config\backslash\config^+_b\cup\config^-_b$. We distinguish the following cases:
	\begin{enumerate}
	\item Let $c=a$, i.e., $\loc(a)\in\config$. 
	By $\indiscernible{a}{b}{\sconfig}$, we also have $\loc(b)\in\config$. 
	Note that $\langle\loc(b),w\rangle\sim_\pi\langle\loc(a),\pi(w)\rangle$. 
	Note that $\loc(b)\neq \loc^-(b)$. %
	By assumption, $\succ_\B(\loc(b),w)$ is non-accepting. 
	By Corollary \ref{corollary:gnra_bad},  $\succ_\B(\loc(a),\pi(w))$ is non-accepting, too. 
	\item Let $c\neq a$. 
	Note that also $\langle\loc^-(b),w\rangle \sim_\pi \langle\loc^-(a),\pi(w)\rangle$. 
	Recall that $\succ_\B(\loc^-(b),w)$ is accepting. 
	By Corollary \ref{corollary:gnra_bad},  $\succ_\B(\loc^-(a),\pi(w))$ is accepting. 
	We prove below that $\{\loc^-(a),\loc(c)\}$ is coverable.
	By unambiguity of $\B$, this directly implies that $\succ_\B(\loc(c),\pi(w))$ is non-accepting.

	 Recall that $\{d\in\N\mid \loc^-(d)\in C\}$ is cofinite. 
	 Pick some datum $d\in\N\backslash\{c\}$ such that $\loc^-(d)\in\config$. We distinguish two cases. 
	 
	 \begin{itemize}
	 \item Assume 
	 $\loc(c)\in\config\backslash\config^+_b$. 
	 Since $\config$ is essentially coverable, 
	 the set $\{\loc^-(d), \loc(c)\}$ is coverable. 
	 Hence there must exist some data word $u$ such that $\{\loc^-(d),\loc(c)\}\subseteq \succ_\B(\loc_\init(\bot),u)$. 
	Let $\pi'$ be a partial isomorphism satisfying $\pi'(d)=a$, $\pi'(a)=d$,  and $\pi'(e)=e$ for all $e\in\data(u)\cup\{c\}$. 
	Then $\{\loc^-(a),\loc(c)\}\subseteq \succ_\B(\loc_\init(\bot),\pi'(u))$, hence $\{\loc^-(a),\loc(c)\}$ is coverable. 

	\item  Second suppose $\loc(c)\in\config^-_b$, i.e., $c=b$. 
	 This implies that $\{e\in\N\mid \loc(e)\in\config\}$ is cofinite. 
	 Pick some datum $e\in\N\backslash\{d\}$ such that $\loc(e)\in\config$. 
	 Since $\config$ is essentially coverable, 
	 the set $\{\loc^-(d),\loc(e)\}$ is coverable. 
	 Hence there must exist some data word $u$ such that $\{\loc^-(d),\loc(e)\}\subseteq\succ_\B(\loc_\init(\bot),u)$. 
	 Let $\pi'$ be a partial isomorphism satisfying $\pi'(d)=a$, $\pi'(a)=d$, $\pi'(b)=e$, $\pi'(e)=b$, and $\pi'(f)=f$ for all $f\in\data(u)$. Then $\{\loc(b),\loc^-(a)\}\subseteq \succ_\B(\loc_\init(\bot),\pi'(u))$, hence $\{\loc(c),\loc^-(a)\}$ is coverable.
	 \end{itemize}
	\end{enumerate}
	Altogether we have proved that $\succ_\B((\config\backslash\config^+_b)\cup\config^-_b,\pi(w))$ is non-accepting, while there is an accepting run of $\A$ on $\pi(w)$ starting in $\loc^\A(\vect{d})$. 
	This finishes the proof for the $(\Rightarrow)$-direction, and thus the proof of the Proposition. 	
\end{proof}

As in~\cite{DBLP:conf/stacs/MottetQ19}, Proposition~\ref{prop:collapse_gura} is enough to obtain an \expspace\ algorithm deciding containment, proving Theorem~\ref{thm_gura}.

\bibliographystyle{plain}
\bibliography{citat}

\end{document}